\documentclass{article} 
\usepackage{iclr2020_conference,times}

\usepackage{amsmath,amsfonts,bm}

\def\eqref#1{equation~\ref{#1}}

\def\1{\bm{1}}

\DeclareMathAlphabet{\mathsfit}{\encodingdefault}{\sfdefault}{m}{sl}
\SetMathAlphabet{\mathsfit}{bold}{\encodingdefault}{\sfdefault}{bx}{n}

\usepackage{hyperref}
\usepackage{url}
\usepackage{graphicx}
\usepackage{wrapfig}

\newcommand{\greencheck}{{\color{green}\checkmark}}

\title{Population Mapping in Informal Settlements with High-Resolution Satellite Imagery and Equitable Ground-Truth}

\author{Konstantin Klemmer \\
University of Warwick \& NYU \thanks{Corresponding author: \texttt{k.klemmer@warwick.ac.uk}} \\
\And
Godwin Yeboah \\
University of Warwick \\
\And
Jo\~{a}o Porto de Albuquerque \\
University of Warwick \& \\ The Alan Turing Institute\\
\And
Stephen A. Jarvis \\
University of Warwick \& \\ The Alan Turing Institute\\
}

\iclrfinalcopy 
\begin{document}

\maketitle

\begin{abstract}
We propose a generalizable framework for the population estimation of dense, informal settlements in low-income urban areas--so called 'slums'--using high-resolution satellite imagery. Precise population estimates are a crucial factor for efficient resource allocations by government authorities and NGO's, for instance in medical emergencies. We utilize \textit{equitable} ground-truth data, which is gathered in collaboration with local communities: Through training and community mapping, the local population contributes their unique domain knowledge, while also maintaining agency over their data. This practice allows us to avoid carrying forward potential biases into the modeling pipeline, which might arise from a less rigorous ground-truthing approach. We contextualize our approach in respect to the ongoing discussion within the machine learning community, aiming to make real-world machine learning applications more inclusive, fair and accountable. Because of the resource intensive ground-truth generation process, our training data is limited. We propose a gridded population estimation model, enabling flexible and customizable spatial resolutions. We test our pipeline on three experimental site in Nigeria, utilizing pre-trained and fine-tune vision networks to overcome data sparsity. Our findings highlight the difficulties of transferring common benchmark models to real-world tasks. We discuss this and propose steps forward.
\end{abstract}

\section{Introduction}\label{Section1}

As urbanization accelerates around the globe, cities increasingly struggle to provide affordable and adequate housing for their growing population. This has forced many people seeking opportunities and employment in the cities into informal settlements, also referred to as 'slums'. These informal settlements are often highly deprived and their populations face various problems, e.g. the lack of healthcare access \citep{VandeVijver2015} or basic infrastructure \citep{Castells-Quintana2017}. The poor living conditions in informal settlements in combination with their continuous growth have led the United Nations (UN) to specifically mention "slums" as a challenge to their sustainable development goals (SDG) \citep{UnitedNations2018}. It is thus crucial to understand informal settlements and their development. In order to optimize humanitarian aid, precise "slum" detection and population estimates are of particular interest. Deep learning approaches built atop satellite image data have recently become more and more popular for this task \citep{Jean2016}, as they scale well on high-dimensional data and outperform traditional methods of remote sensing  \citep{Persello2017}. However, these approaches mostly originate from the remote sensing, geographic information science (GIS) and human geography communities \citep{Kuffer2016}. We believe that the machine learning community should assume a more active role tackling these problems, especially since it is at the forefront of computer vision, pattern recognition and scalable deep learning methods. But while image classification is one of the most successful applications of machine learning, it often struggles to generalise to real-life cases such as multi-label images and poor-resolution images. In remote sensing and satellite image classification, we see this effect demonstrated in competitions such as Deep Globe \cite{Demir2018} in which building detection for Las Vegas was 85\% whereas detection for Khartoum was only 54\%. This also highlights another problem: current methods work better on cities in the Western world, while cities in the Global South, which may benefit especially from these applications, remain neglected. 

In this study, we propose a flexible framework for estimating the population of informal settlements. Our proposed method builds on high-resolution satellite imagery and utilizes equitable ground-truth: Building-level shapes and labels are mapped and curated in collaboration with local universities and communities. Through this cooperative approach, the local population takes agency over mapping their own neighborhood; furthermore leveraging their unique domain knowledge to generate high-quality data. While satellite images have previously been used to map informal settlements, our equitable ground-truth data presents a novel approach to population mapping. We further comment on the differences with existing population mapping approaches. Our proposed pipeline is based on a gridded population estimation model with adaptable spatial resolution. Methodologically, we have to overcome a lack of training data, stemming from the expensive ground-truth generation process, and resulting proneness to overfitting. We experiment with fine-tuned versions of pre-trained computer vision algorithms like \textit{ResNet50} and \textit{MobileNetV2} to tackle this issue. However, we find a surprising lack of capability of these powerful networks. We use these preliminary finding as a starting point to discuss the challenges of successfully implementing our approach in terms of (1) the ethical and qualitative data requirements and (2) the methodological requirements in the light of inadequate, existing benchmark models. Lastly, we outline a research agenda to address these issues.

\section{Related Work}\label{Section2}

\begin{table}[]
\centering
\scalebox{0.8}{
    \begin{tabular}{l|llllll}
     Publication & Data (input) & Data (labels) & Output res. & Valid. res. & Method & Eq. GT \\ \hline
     \hline
     \citealt{Wardrop2018} & census data & - & \textit{flex. grid} & - & Spatial &  - \\
     & soc.-ec. covariates &  &  &  & Disaggregation &  \\
     & built environment &  &  &  &  & \\
     \hline
     \citealt{Zong2019} & coarse labels & cellular data & \textit{flex. grid} & cell-tower & Deep &  - \\
     & POI data & & & & Learning & \\
     \hline
     \citealt{Weber2018} & high-res & survey data & \textit{flex. grid} & - & Image &  - \\
     & satellite image & & & & Segmentation & \\
     & microcensus & & & & & \\
     \hline
     \citealt{Hu2019} & high-res & census data & \textit{flex. grid} & village & Deep &  - \\
     & satellite image & & & & Learning & \\
     \hline
     \citealt{Xie2015} & nighttime & poverty index & \textit{flex. grid} & household & Deep &  - \\
     & light & & & group & Learning & \\
     & satellite image & & & & & \\
     \hline
     \textbf{Ours} & high-res & building mask & \textit{flex. grid} & building & Deep & \greencheck \\
     & satellite image & & & & Learning & 
    \end{tabular}}
    \caption{Comparing recent population estimation methods: the \textit{Data} columns describe input and validation labels (if present), \textit{Output res.} and \textit{Validation res.} provide the spatial units of the estimation output and the validation labels respectively, \textit{Method} gives the quantitative approach for population estimation and \textit{Eq. GT} indicates, if the validation labels constitute equitable ground-truth.}
    \label{table:1}
\end{table}

Informal settlement mapping is an increasingly prominent research objective \citep{Hofmann2015}, but poses distinct challenges, as the morphology and structure of informal settlements varies substantially--across different cultural spaces and even within cities \citep{Taubenbock2018}. As of recent, machine learning approaches provide promising solutions to account for these complexities \citep{Maiya2018,Wurm2019}. Yet, few studies are built around active community engagement, such as participatory mapping of informal settlements \citep{Yuan2018}. We believe integrating local communities into work such as ours is not only helpful for leveraging the locals' expertise, but also provides them with agency over their data, enables further conversation on the equitability and performance of resulting applications and prevents model bias stemming from poor-quality ground-truth. Informal settlements can be seen as self-contained systems, including their own places of worship, entertainment and trade. Many structures are not inhabited, but rather being used for other purposes (e.g. shops or religious buildings). Precise population estimates are thus often hard to obtain. Recently, Wardrop et al (\citeyear{Wardrop2018}) highlighted the need for better and more granular population estimates in informal settings. Traditionally, spatial disaggregation approaches utilizing auxiliary information such as road networks or vegetation have been used \citep{Wardrop2018}, while more recent machine learning focused work has proposed disaggregation using points-of-interest \citep{Zong2019} or combining satellite images with existing census data \citep{Weber2018,Hu2019}. A further, methodological challenge is that few deep learning methods tailored to satellite imagery exist. Together with the fact that applicable training data and reliable ground-truth can be sparse and expensive to produce, building models which can be trained on limited data but perform well on a global scale are the "holy grail" of informal settlement analysis. So far, this task has been proven to be exceptionally challenging: Wurm et al. (\citeyear{Wurm2019}) point out that data quality and resolution differences are a substantial problem in their transfer learning task on informal settlement satellite imagery. However, Xie et al. (\citeyear{Xie2015}) find that rather simple data such as nighttime lights can be used for successful transfer learning and poverty mapping. They also note that pre-training on standard datasets such as \textit{ImageNet} can be useful. Table 1 provides an overview on data requirements, spatial resolution, methods and equitability of existing approaches, as compared to our proposed approach.

\section{Data \& Equitable ground-truth}\label{Section3}

\begin{figure}
\centering
\includegraphics[scale=0.5]{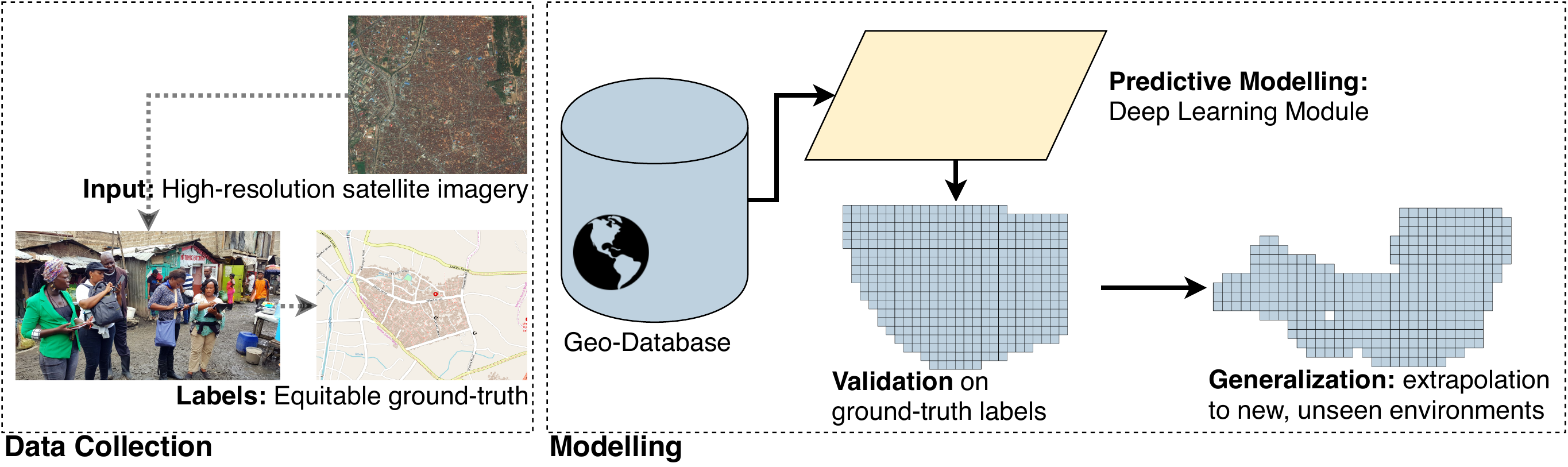}
\caption{\label{fig:fig1}Proposed population estimation framework with equitable ground-truth through participatory online digital mapping with on-site validation; Figure partially adapted from Porto de Albuquerque et al. (\citeyear{PortoDeAlbuquerque2019}).}
\vspace{-4mm}
\end{figure}

Our experiments are conducted using data from three sites: two settlements in Ibadan, Nigeria (Sasa, Idikan) provide training data and one in Lagos, Nigeria (Bariga) test data. Our data can broadly be split into two categories: Satellite image data (\textit{input}) and ground-truth maps (\textit{labels}): 

\textbf{Input}: We acquire very high-resolution (VHR) satellite image data with RGB channels and a pixel length of 0.02 meters for each of the three test sites from two major mapping services via the \textit{SAS Planet} software \footnote{Accessible here: \url{http://www.sasgis.org/}}, obtaining two images per informal settlement. Nevertheless, this sub-meter resolution is not indicative of quality; rather we find the images to be quite noisy, especially compared to VHR images of high-income cities such as New York. Satellite data was collected in 2019, the ground-truth labels were mapped in 2018 (note that the images used to create the ground-truth-maps are different from the images we use for model training). We find that the temporal difference in accessing the respective data does not affect the match between both datasets. This implies that all three informal settlements did not change substantially during the period between community-mapping and accessing satellite data. Nevertheless, informal settlements are inherently dynamic and if different imagery is used for mapping and modeling, the overlap between them should always be assessed to guarantee consistent results. 

\textbf{Labels}: Active community engagement is essential for avoiding biased or misinformed labels which might transfer into biased models. We use data from the University of Warwick's \textit{Improving Health in Slums Collaborative} (IHSC, \citeyear{IHSC2019}), which is derived from a participatory community mapping framework that aims to achieve equitable, empowering community engagement while at the same time conforming to rigorous, reliable and generalizable data standards \citep{PortoDeAlbuquerque2019}. This framework is based on the premise that the generation of data by citizens can also be an opportunity for mutual learning and to develop a new critical consciousness on the local realities that are being mapped in the data \citep{AlbuquerqueAlmeida2020}.  While the \textit{IHSC} paper should be consulted for details, we want to outline the key steps of this approach briefly (also, see Figure 1): In preparation for the mapping, local stakeholders are consulted to identify key local partners. These partners are essential for understanding the complexities of the physical and social environments in informal settlements. The next step involves the creation of a base map from high-resolution satellite imagery. This mapping process can be conducted by locals and remote participants using open-source collaborative software such as the \textit{HOT Tasking Manager} (see: \url{https://tasks.hotosm.org/}). Lastly, "ground-truthing" is conducted at the site using participatory mapping by specially trained local citizens. This validates and enriches the data gathered in the previous step. The final map can then be contributed as open geographic data to volunteered geographic information (VGI) services like \textit{OpenStreetMap}. OpenStreetMap is a "wikipedia of maps"; a collaborative project aim at creating a free editable map of the world. Consequently, our definition of "equitable ground-truth" (and how we compare it to existing work, as in Table \ref{table:1} entails all the principles outlined above: \textit{pre-assessment \& local partners}, \textit{community participation \& education}, \textit{double validation (remote and on-site)} and \textit{transparency \& accessibility}.

\section{Discussion}\label{Section4}


In order to explore how we may learn differently with this community data, we decided to look at the process of classifying buildings into inhabited vs non-inhabited from satellite images. This is a difficult but important question: Building an understanding of where people live is crucial for estimating needs in emergency scenarios and for general public service provision and planning. Below we give some of the steps and results of our preliminary analysis.

 
First we split the satellite images into a grid of 36x36 meter tiles. We then rescale each image from 1600x1600 pixels to 224x224 pixels - given how noisy the images are there was not much loss of information with this compression. For one of the settlements--Lagos-Bariga--we also removed images with cloud cover, calculating the ratio of white pixels as a proxy - we set threshold at 5\%, which gives us as many images as possible. After creating the grid, each image is assigned a respective residential occupancy level (see Figure 1), calculated as the percentage of pixels occupied by residential buildings. We transform this into a classification problem by declaring images with 30\% or more residential occupancy level as residential ($=1$) and less than that as non-residential ($=0$). We chose this threshold through a preceding analysis, comparing each image to its respective building masks and assessing visual cues in the images, while avoiding further class imbalances. Even though this might seem a fairly low residential occupancy level, this threshold corresponds to the true ratio of occupied versus non-occupied buildings, which we observe to be very similar in all three sites. About $1/3$ of the data are assigned a "residential" label and $2/3$ a "non-residential" label.


\begin{figure}
\centering
\includegraphics[scale=0.35]{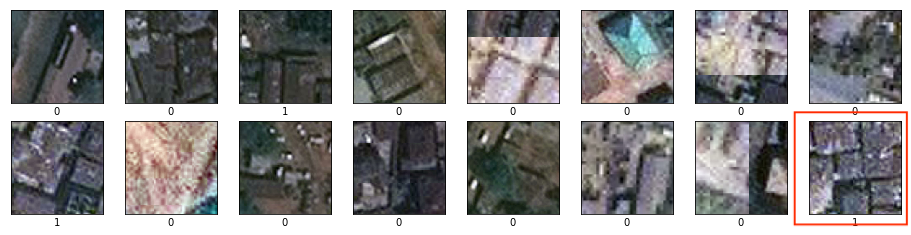}
\caption{\label{fig:test} Random sample of training data. Highlighted on the lower right is an exemplary image for high residency}
\vspace{-4mm}
\end{figure}


For our baseline analysis, we trained a \textit{ResNet50} model (150 epochs) \citep{He2016} and a \textit{MobileNetV2} (30 epochs) \citep{Sandler2018} using data from Ibadan-Sasa and Ibadan-Idikan for training- which gave us a total of 1056 training examples - and from Lagos-Bariga for validation - 646 validation examples. We found that \textit{MobileNetV2}, trained with only our dataset performed better than both \textit{MobileNetV2} with pre-trained weights from the \textit{Imagenet} dataset.  Validation performance converged after running 30/50 epochs. In this set up we got a training accuracy of 75\% and validation accuracy of 63\%. In several studies in machine learning, transfer learning has proven to be a great tool to learn under constrained resources \cite{Zoph_2016,alex2019large}. In our case we find that in questions as complex as that of building separability, standard methods of pre-training with benchmark datasets does not help us much and in fact harms accuracy. This is not a novel finding \cite{ngiam2018domain}, but it once again underlines the importance of curating relevant datasets, especially for applied work. While our preliminary results may suggest that we are not close to a practicable solution, we gained crucial insights to guide future research directions. Further they allow us to comment on lessons-learned and best-practices when working with equitable ground-truth data:

\begin{itemize}
  \item During the data preparation, we realised that there were spatial cues in images such as how square a roof is or how square neighbouring roofs are, that help the classification task. However, we have learned automating this separation is a lot more complex. That said, this might just be a function of the limited dataset.
  \item Ethical considerations regarding the detachment between the "analysts" (e.g. researchers) and "subjects" (informal settlement population) need to take a more central role in machine learning based population mapping tasks. Although different organisations are increasingly discussing issues around bias and adopting frameworks based on principles for ``responsible AI'', there is still a lack of clear processes  for engaging with the local communities that are represented in the data used in machine learning systems.  We hope that by having the local community as part of the annotators, we have provided the community with hitherto unavailable data to understand their contexts, while simultaneously generating a more representative training dataset. However, we also conclude that deeper analyses of the informal and often undocumented implications of machine learning systems are needed. Failing to do so would not only exclude or misrepresent large population parts \footnote{In Nigeria, for example, the informal sector accounts for 50-65\% of GDP. IMF 2017 working paper: \url{https://www.imf.org/en/Publications/WP/Issues/2017/07/10/The-Informal-Economy-in-Sub-Saharan-Africa-Size-and-Determinants-45017}}, but also may lead to severe problems with downstream tasks. It is hence crucial that the machine learning community actively works with practitioners and local stakeholders. Further, best-practice guidelines for generating and working with equitable ground-truth data should be developed, both in general and domain-specific, such as the recent proposal of an integrated deprived area mapping system \citep{Kuffer2020}.
  \item We observe a certain lack of transferability of existing, pre-trained vision models to our task. In particular, we see that larger--and presumably more powerful--models perform worse than smaller models. This raises the question of how much real-world relevance existing benchmarks hold. This also serves as a call-to-action for developing tailored methodologies and domain-specific benchmarks for common applications of machine learning in the development context.
\end{itemize}

\section{Conclusion}\label{Section5}

Government authorities and NGOs require precise population estimates for informal settlements to optimize their services and health interventions. Often the subject of these estimates, the population, is excluded from the data gathering process. In this paper we discuss the merits and difficulties of deploying machine learning for population estimation in informal settlements, building on equitable ground-truth data, gathered in collaboration with local population and stakeholders. This approach aims at empowering local communities to take agency over their own data and at eliminating potential root causes of biases further down the modelling pipeline. We argue that this method is not only more ethical, but also leads to a higher data quality, as local knowledge is explicitly leveraged throughout the annotation process.

We run experiments using pre-trained and fine-tuned vision models. Interestingly, we observe that pre-trained models don't fare as well as may be expected, and smaller models such as \textit{MobileNetV2} perform best. This raises concerns on the value ascribed to common computer vision benchmark models and their transferability to real world tasks. We hope our findings encourage further investigations in this area within the ML for development community.

\subsubsection*{Acknowledgments}
First, we wish to that thank Nyalleng Moorosi for her guidance and contribution to this project. The authors also gratefully acknowledge funding from the UK Engineering and Physical Sciences Research Council, the EPSRC Centre for Doctoral Training in Urban Science (EPSRC grant no. EP/L016400/1); The Alan Turing Institute (EPSRC grant no. EP/N510129/1). The authors would like to express gratitude to all project members of the NIHR Improving Health in Slums Collaborative for their invaluable input, to informal settlement residents and to the mappers of the OpenStreetMap community who were active locally and remotely to create the data used in this paper.

\bibliography{iclr20}
\bibliographystyle{iclr2020_conference}


\end{document}